\documentstyle[12pt]{JHEP3}

\newcommand{\e}{\epsilon}

\newcommand{\be}[1]{ \begin{equation}\label{#1} }
\newcommand{\ee}{\end{equation}}
\newcommand{\ben}[1]{\begin{eqnarray}\label{#1} }
\newcommand{\een}{\end{eqnarray}}
\newcommand{\eq}[1]{(\ref{#1})}
\def\ZZZ{{\hskip-3pt\hbox{ Z\kern-1.6mm Z}}}
\def\zzz{{\hskip-3pt\hbox{ z\kern-1mm z}}}

\newcommand{\p}{\partial}

\newcommand{\refb}[1]{(\ref{#1})}

\def\one{{\hbox{ 1\kern-.8mm l}}}
\def\zero{{\hbox{ 0\kern-1.5mm 0}}}

\title{Galilean Conformal Algebras and AdS/CFT}

\author{
Arjun Bagchi, Rajesh Gopakumar \\
$\;$ Harish-Chandra Research Institute,\\
$\;$ Chhatnag Road,\\
$\;$ Jhusi. India 211019.\\
$\;$\email{arjun, gopakumr@hri.res.in}
}

\abstract{Non-relativistic versions of the AdS/CFT conjecture have recently been investigated in some detail.
These have primarily been in the context of the Schrodinger symmetry group. Here we initiate a study
based on a {\it different} non-relativistic conformal symmetry: one obtained by a parametric 
contraction of the relativistic conformal group. The resulting Galilean conformal symmetry has the same number of generators as the relativistic symmetry group and thus is different from the Schrodinger group (which has fewer). One of the interesting features of the Galilean Conformal Algebra is that it admits an extension to an {\it infinite} dimensional symmetry algebra (which can potentially be dynamically realised). 
The latter  contains a Virasoro-Kac-Moody 
subalgebra. We comment on realisations of this extended symmetry in a boundary field theory. We also propose a somewhat unusual geometric structure for the bulk gravity dual to any realisation of this symmetry. This involves taking a Newton-Cartan like limit of Einstein's equations in anti de Sitter space which singles out an $AdS_2$ comprising of the time and radial direction. 
The infinite dimensional Virasoro extension is identified with the asymptotic isometries of this $AdS_2$.}

\begin{document}

\baselineskip 3.5ex




\section{Introduction}
Even after more than a decade, the AdS-CFT conjecture \cite{Maldacena:1997re} continues to throw
up rich, new avenues of investigation. One such recent direction has been to consider extensions of the 
conjecture from its original relativistic setting to a non-relativistic context. This opens the door to potential applications of the spirit of gauge-gravity duality to a  variety of real-life strongly interacting systems. 
It was pointed out in \cite{Nishida:2007pj} that the Schrodinger symmetry group \cite{Hagen:1972pd, Niederer:1972zz}, a non-relativistic version of conformal symmetry, is relevant to the study of cold atoms. A gravity dual possessing these symmetries was then proposed in \cite{Son:2008ye, Balasubramanian:2008dm} (see also \cite{Goldberger:2008vg, Barbon:2008bg} for a somewhat different bulk realisation).
Further developments along this line can be found in  \cite{Herzog:2008wg} --\cite{{Akhavan:2009ns}}. 

Instead of the Schrodinger group, in this paper, we will consider an {\it alternative} non-relativistic 
realisation of conformal symmetry and begin a study of its consequences and realisations in the context of the AdS/CFT conjecture. This symmetry will be obtained by considering the nonrelativistic group contraction of the relativistic conformal group $SO(d+1,2)$ in $d+1$ space-time dimensions. \footnote{The process of group contraction is, of course, standard and may have been applied by many people to the relativistic conformal group. To the best of our knowledge, \cite{Lukierski:2005xy} is a recent reference with the explicit results of this contraction of the relativistic conformal group. This reference goes on to study a realisation of the $2+1$ dimensional case, which has some special features.} The process of group contraction leads, in $d=3$, for instance, to a fifteen parameter group (like the parent $SO(4,2)$ group) which contains the ten parameter Galilean subgroup. This Galilean conformal group is to be contrasted with the twelve parameter Schrodinger group (plus central extension) with which it has in common only the Galilean subgroup. The Galilean conformal group is, in fact, different from the Schrodinger group in some crucial respects, which we will describe in more detail later.  For instance, 
the dilatation generator $\tilde{D}$  in the Schrodinger group scales space and time differently 
$x_i \rightarrow \lambda x_i, t\rightarrow \lambda^2 t$. Whereas the corresponding generator $D$ in the Galilean Conformal Algebra (GCA)
scales space and time in the {\it same} way $x_i \rightarrow \lambda x_i, t\rightarrow \lambda t$. Relatedly, the GCA does {\it not} admit a mass term as a central extension. Thus, in some sense, this
symmetry describes "massless" or "gapless" non-relativistic theories, like the parent relativistic group but unlike the Schrodinger group.   

However, the most interesting feature of the GCA seems to be its natural extension to an
infinite dimensional symmetry algebra (which we will also often denote as GCA when there is no risk of confusion). This is somewhat analogous, as we will see, to the way in which the finite conformal algebra of $SL(2,C)$ in two dimensions extends to two copies of the Virasoro algebra. We will see that it is natural to expect this extended symmetry to be dynamically realised (perhaps partially) in actual systems possesing the finite dimensional Galilean conformal symmetry.
 Indeed, it has been known (see \cite{Duval:1993pe} and references therein) that there is a notion of a "Galilean isometry"  which encompasses the so-called Coriolis group of arbitrary time dependent (but spatially homogeneous) rotations and translations.  In this language, our infinite dimensional algebra is that of "Galilean conformal isometries".  As we will see, it contains one copy of a Virasoro  together with an $SO(d)$ current algebra (on adding the appropriate central extension).  

In addition to possible applications to non-relativistic systems, one of the motivations for studying the contracted $SO(d+1,2)$ conformal algebra is to examine the possibility of a new tractable limit of the parent AdS/CFT conjecture. In fact, the BMN limit \cite{BMN} of the AdS/CFT conjecture is an instance where, as result of taking a particular scaling limit,  one obtains a contraction of the original $SO(4,2)\times SO(6)$ (bosonic) global symmetry \footnote{An algebraically equivalent contraction to ours, of the isometries of $AdS_5\times S^5$, was studied in \cite{Gomis:2005pg} as an example of a non-relativistic string theory. However, the embedding of this contraction in $AdS_5$ is not manifestly such that it corresponds to a  non-relativistic CFT on the boundary. We will comment further on this at a later stage.}.  In our case, the non-relativistic contraction is obtained by taking a similar scaling limit on the parent theory. Like in the BMN case, taking this limit 
would isolate a closed subsector of the full theory. The presence of an enhanced symmetry in our scaling limit raises interesting possibilities about the solvability of this subsector. We defer the detailed study of this aspect for later. 

There are, however, some important differences here from a BMN type limit which have to do with the nature of taking the scaling. Normally the BMN type scaling leads to a Penrose limit of the geometry in the vicinity of some null geodesic.  These are typically pp-wave 
like geometries whose isometry is the same as that of the contracted symmetry group on the boundary. 
The non-relativistic scaling limit that leads to the GCA on the boundary is at first sight more puzzling 
to implement in the bulk. This is because, under  the corresponding scaling, the bulk metric degenerates in the spatial directions $x_i$. Thus one might think one has some kind of singular limit in the bulk 
description. However, this degeneration is a feature common to all non-relativistic limits. It arises, for instance, in taking the Newtonian gravity limit of Einstein's equations in asymptotically flat space. 
However, in this case, there is a well defined geometric description of the limit despite the degeneration of the metric. This description, originally due to Cartan, and studied fairly thoroughly by geometers describes Newtonian gravity in terms of a non-dynamical metric but with a dynamical {\it non-metric}  connection\footnote{See the textbook \cite{Misner:1974qy} Chap.12 for a basic discussion. We thank T. Padmanabhan for bringing this to our attention.}. 
The Einstein's equations reduce to equations determining the curvature of this connection 
in terms of the matter density. These are nothing but the Poisson equations for the Newtonian gravitational potential. In geometric terms the spacetime takes the form of a vector bundle with fibres as the spatial $R^{d}$ over a base $R$ which is time, together with an affine connection related to the gradient of the Newtonian potential. 
 
We propose a similar limiting description for the bulk geometry in our case. The main difference is that
the time  and the radial direction together constitute an $AdS_2$ with a non-degenerate metric. Thus one has a geometry with an $AdS_2$ base and the spatial $R^{d}$  fibred over it. Once again there is no overall spacetime metric. The dynamical variables are  affine connections determined by the limiting form of Einstein's equations.  As a check of this proposal, we will see that the 
infinite dimensional GCA symmetries are realised in this bulk geometry as asymptotic isometries.
In fact the Virasoro generators of the GCA are precisely the familiar generators of asymptotic global 
isometries of $AdS_2$. These generators will also be seen to reduce to the generators of the GCA on the boundary. 

The paper is organised as follows. In the next section we first review the Schrodinger symmetry algebra
in order to set notation and contrast it with the Galilean Conformal Algebra, which we obtain through
group contraction on $SO(d+1,2)$. In Sec. 3 we describe the infinite dimensional extension of this algebra and its physical significance. Sec. 4 moves onto the bulk realisation of the non-relativistic contraction. 
We propose a geometric description of the bulk physics analogous to the Newton-Cartan theory.
In Sec. 5 we lend support to this proposal by finding the vector fields corresponding to the GCA and its infinite dimensional extension as well as their action on the bulk $AdS$. In Sec. 6 we close with a laundry list of things left undone. In two appendices we elaborate on a couple of points of the main text.

\section{Non-Relativistic Conformal Symmetries}

\subsection{Schrodinger Symmetry}

The Schrodinger symmetry group in $(d+1)$ dimensional spacetime (which we will denote as 
$Sch(d,1)$) has been studied as a non-relativistic analogue of conformal symmetry. It's name 
arises from being the group of symmetries of the free Schrodinger wave operator in $(d+1)$ 
dimensions. In other words, it is generated by those transformations that commute with the operator 
$S=i\p_t+{1\over 2m}\p_i^2$. However, this symmetry is also believed to be realised in 
interacting systems, most recently in cold atoms at criticality. 

The symmetry group contains the usual Galilean group (denoted as $G(d,1)$) with its central 
extension. 
\ben{galalg}
[ J_{ij}, J_{rs} ] &=& so(d) \cr
[J_{ij} , B_r ] &=& -(B_i {\delta}_{jr} - B_j {\delta}_{ir}) \cr
[J_{ij},\, P_r] &=& -(P_i {\delta}_{jr} - P_j {\delta}_{ir}), \quad [J_{ij},\, H] = 0 \cr
 [B_i,B_j] &=& 0, \quad [P_i, P_j] =0 ,\quad [B_i, P_j] =m\delta_{ij} \cr
[H, P_i] &=& 0, \quad [H, B_i] = - P_i. 
\een
Here $J_{ij}$ $(i,j=1\ldots d)$ are the usual $SO(d)$ generators of spatial rotations. $P_r$ are the $d$ generators of spatial translations and $B_j$ those of boosts in these directions. Finally $H$ is the generator of time translations. The parameter $m$ is the central extension and has the interpretation
as the non-relativistic mass (which also appears in the Schrodinger operator $S$).

As vector fields on the Galilean spacetime $R^{d,1}$, they have the realisation (in the absence of the
central term) 
\ben{galvec}
J_{ij} &=& -(x_i\p_j-x_j\p_i) \qquad H= -\p_t \cr
P_i &=& \p_i \qquad B_i= t\p_i 
\een
 
In addition to these Galilean generators there are {\it two} more generators which we will denote 
by $\tilde{K}, \tilde{D}$.  
$\tilde{D}$ is a dilatation operator, which unlike the relativistic case, scales time and 
space differently. As a vector field $\tilde{D}= -(2t\p_t+x_i\p_i)$ so that
\be{tildact}
x_i \rightarrow \lambda x_i ,  \qquad t \rightarrow \lambda^2 t.
\ee
$\tilde{K}$ acts something like the time component of special conformal transformations. 
It has the form $\tilde{K} = -(tx_i\p_i+t^2\p_t)$ and generates the finite transformations 
(parametrised by $\mu$)
\be{tilkact}
x_i \rightarrow {x_i\over (1+\mu t)} ,\qquad  t \rightarrow {t \over (1+\mu t)}.
\ee

These two additional generators have non-zero commutators
\ben{schaddl}
[\tilde{K}, P_i] &=& B_i, \quad [\tilde{K}, B_i] = 0,  \quad [\tilde{D} , B_i]=-B_i  \cr 
[\tilde{D} , \tilde{K}] &=& -2\tilde{K}, \quad [\tilde{K}, H]= -\tilde{D}, \quad [\tilde{D}, H] = 2H.
\een
The generators $\tilde{K}, \tilde{D}$ are invariant under the spatial rotations $J_{ij}$.
We also see from the last line that $H, \tilde{K}, \tilde{D}$ together form an $SL(2,R)$ algebra. 
The central extension term of the Galilean algebra is compatible with all the 
extra commutation relations. 

Note that there is no analogue in the Schrodinger algebra of the spatial components $K_i$ of special conformal transformations. 
Thus we have a  smaller group compared to the relativistic conformal group. In $(3+1)$ dimensions
the Schrodinger algebra has twelve generators (ten being those of the Galilean algebra) and the additional central  term. Whereas the relativistic conformal group has fifteen generators. 
In the next subsection we will discuss how to get a nonrelativistic conformal group through group contraction. In the process of group contraction one does not lose any generators and hence the Galilean Conformal Algebra we find will have the same number of generators as the group
$SO(d+1,2)$.

\subsection{Contraction of the Relativistic Conformal Group}

We know that the Galilean algebra $G(d,1)$ arises as a contraction of the Poincare algebra
$ISO(d,1)$. Physically this comes from taking the non-relativistic scaling 
\be{nrelscal}
t \rightarrow \epsilon^r t \qquad   x_i \rightarrow \epsilon^{r+1} x_i
\ee
with $\epsilon \rightarrow 0$. This is equivalent to taking the velocities $v_i \sim \epsilon$ to zero
(in units where $c=1$). We have allowed for a certain freedom of scaling through the parameter $r$,  since we might have other scales in the theory with respect to which we would have to take the above nonrelativistic limit. We will later consider the example of nonrelativistic fluid mechanics, in which we have a scale set by the temperature. In this case the natural scaling corresponds to $r=-2$. However, for the process of group contraction the parameter $r$ will play no role apart from modifying an over all factor which is unimportant. Hence we will mostly take $r=0$. 

Starting with the expressions for the Poincare generators ($\mu,\nu=0,1\ldots d$)
\be{}
J_{\mu\nu} = -(x_{\mu} \p_{\nu} - x_{\nu} \p_{\mu}) \qquad P_{\mu}=\p_{\mu},
\ee
the above scaling gives us the Galilean vector field generators of \eq{galvec}
\begin{eqnarray}
J_{ij}&=& -(x_i\p_j-x_j\p_i) \qquad P_0=H= -\p_t \cr
P_i &=& \p_i \qquad J_{0i}=B_i= t\p_i .
\end{eqnarray}

They obey the commutation relations (without central extension) of \eq{galalg}.
This should be contrasted with the Poincare commutators
\ben{poinalg}
[  J_{ij},  J_{rs} ] &=& so(d) \cr
[ J_{ij} ,  B_r ] &=& -( B_i {\delta}_{jr} - B_j {\delta}_{ir}) \cr
[ J_{ij},\, P_r] &=& -(P_i {\delta}_{jr} - P_j {\delta}_{ir}), \quad [J_{ij}, H] = 0 \cr
[B_i,B_j] &=& -J_{ij}, \quad [ P_i,  P_j] =0 ,
\quad [ B_i,  P_j] =\delta_{ij} H\cr
[ H,  P_i] &=& 0, \quad [ H,  B_i] =  -P_i  
\een

To obtain the Galilean Conformal Algebra, we simply extend the scaling \eq{nrelscal} to the rest of the generators of the conformal group $SO(d+1,2)$. Namely to 
\be{}
D = -(x\cdot\p) \qquad K_{\mu} = -(2x_{\mu}(x\cdot\p) -(x\cdot x)\p_{\mu})
\ee
where $D$ is the relativistic dilatation generator and $K_{\mu}$ are those of special conformal transformations.
The non-relativistic scaling in \eq{nrelscal} now gives (see also \cite{Lukierski:2005xy})
\ben{nrelconf}
D &=& -( x_i \p_i + t \p_t) \cr
K&=& K_0= -(2t x_i \p_i +t^2 \p_t) \cr
K_i &=& t^2\p_i. 
\een

Note that the dilatation generator $D=-( x_i \p_i + t \p_t)$ 
is the {\it same} as in the relativistic theory. It scales space and time in the same way $x_i \rightarrow \lambda x_i , t \rightarrow \lambda t $. Therefore it is different from the dilatation generator 
$\tilde{D} =-(2t\p_t+x_i\p_i)$ of the Schrodinger group. 
Similarly, the temporal special conformal generator $K$ in \eq{nrelconf} is 
different from $\tilde{K} = -(tx_i\p_i+t^2\p_t)$.
Finally, we now have spatial special conformal transformations $K_i$ which were not present in the Schrodinger algebra. Thus the generators of the Galilean Conformal Algebra are 
$(J_{ij}, P_i, H, B_i, D, K, K_i)$.

Since the usual Galilean algebra $G(d,1)$ for the generators $(J_{ij}, P_i, H, B_i)$ is a 
subalgebra of the GCA, we will not write down their commutators. The other non-trivial commutators 
of the GCA are \cite{Lukierski:2005xy}
\ben{galconalg}
[K, K_i] &=&0, \quad [K, B_i]=K_i, \quad [K, P_i]= 2B_i \cr
[J_{ij}, K_r] &=& -(K_i {\delta}_{jr} - K_j {\delta}_{ir}), \quad [J_{ij}, K] =0, \quad [J_{ij}, D]=0 \cr
[K_i,K_j] &=& 0, \quad [K_i, B_j]=0, \quad [K_i,P_j]=0, \quad [H, K_i] = -2B_i, \cr
[D, K_i] &=& -K_i, \quad [D , B_i]=0, \quad [D, P_i] = P_i,\cr
[D,H] &=& H,  \quad [H, K]= -2D, \quad [D, K]=-K.
\een

This can again be contrasted with commutators of the corresponding relativistic generators
\ben{poinconalg}
[K, K_i] &=&0, \quad [ K,  B_i]= K_i, \quad [ K,  P_i]= 2 B_i \cr
[ J_{ij},  K_r] &=& -( K_i {\delta}_{jr} - K_j {\delta}_{ir}), 
\quad [ J_{ij},  K] =0, \quad [J_{ij},  D]=0 \cr
[K_i, K_j] &=& 0, \quad [ K_i,  B_j]= \delta_{ij} K, \quad [ K_i, P_j]
=2J_{ij} +2\delta_{ij}D \cr 
[ H,  K_i] &=& -2 B_i, \quad [ D,  K_i] = - K_i, 
\quad [ D ,  B_i]=0, \quad [D,  P_i] =  P_i,\cr
[D, H] &=& H,  \quad [ H, K]= -2D, \quad [ D, K]=- K.
\een

We can also compare the relevant commutators in \eq{galconalg} with those of 
\eq{schaddl} and we notice that they too are different. Thus the Schrodinger algebra and the 
GCA only share a common Galilean subgroup and are otherwise different. 
In fact, one can verify using the Jacobi identities for $(D,B_i, P_j)$ that the Galilean 
central extension in $[B_i,P_j]$ is {\it not} admissible in the GCA. This is another difference from
the Schrodinger algebra, which as mentioned above, does allow for the central extension. 
Thus in some sense, the GCA is the symmetry  of a "massless" (or gapless) nonrelativistic system. 
We will discuss some possible realisations in the next section. 
It should be pointed out that the GCA does admit a {\it different} central extension of the form
\be{galconcent}
[K_i,P_j]=N\delta_{ij}
\ee
where $N$ commutes with all the other generators of the GCA. The exact interpretation of this term in general is not clear.
 It will, in fact, turn out to be absent when one considers the infinite dimensional extension of the GCA
in the next section.

\section{The Infinite Dimensional Extended GCA}

The most interesting feature of the GCA is that it admits a very natural extension to an infinite 
dimensional algebra of the Virasoro-Kac-Moody type\footnote{After obtaining these results we came to learn of a similar Virasoro  extension of the Schrodinger group\cite{Henkel:1993sg}. The actual algebra is different from the one described here.}. To see this we denote 
\ben{rename}
L^{(-1)} &=& H, \qquad L^{(0)}=D, \qquad L^{(+1)}= K, \cr
M_i^{(-1)} &=& P_i, \qquad M_i^{(0)}=B_i, \qquad M_i^{(+1)}=K_i.
\een
The finite dimensional GCA which we had in the previous section can now be recast as
\ben{gcafinit}
[J_{ij}, L^{(n)}] &=&0 ,  \qquad [L^{(m)}, M_i^{(n)}] =(m-n)M_i^{(m+n)} \cr
[J_{ij} , M_k^{(m)} ] &=& -(M_i^{(m)} {\delta}_{jk} - M_j^{(m)} {\delta}_{ik}), 
\qquad [M_i^{(m)}, M_j^{(n)}] =0, \cr
[L^{(m)}, L^{(n)}] &=& (m-n)L^{(m+n)}.
\een
The indices $m,n=0,\pm 1$
We have made manifest the $SL(2,R)$ subalgebra with the generators $L^{(0)}, L^{(\pm 1)}$. 
In fact, we can define the vector fields 
\ben{gcavec}
L^{(n)} &=& -(n+1)t^nx_i\p_i -t^{n+1}\p_t \cr
M_i^{(n)} &=& t^{n+1}\p_i 
\een 
with $n=0,\pm 1$. These (together with $J_{ij}$) are then exactly the vector fields
in \eq{galvec} and \eq{nrelconf} which generate the GCA (without central extension). 

If we now consider the vector fields of \eq{gcavec} for {\it arbitrary} integer $n$, and also define
\be{Jn}
J_a^{(n)} \equiv J_{ij}^{(n)}= -t^n(x_i\p_j-x_j\p_i)
\ee
then we find that this collection obeys the current algebra 
\ben{vkmalg}
[L^{(m)}, L^{(n)}] &=& (m-n)L^{(m+n)} \qquad [L^{(m)}, J_{a}^{(n)}] = -n J_{a}^{(m+n)} \cr
[J_a^{(n)}, J_b^{(m)}]&=& f_{abc}J_c^{(n+m)} \qquad  [L^{(m)}, M_i^{(n)}] =(m-n)M_i^{(m+n)}. 
\een
The index $a$ labels the generators of the spatial rotation group $SO(d)$ and $f_{abc}$ are the
corresponding structure constants. 
We see that the vector fields generate a $SO(d)$ Kac-Moody algebra without any central terms. In addition to the Virasoro and current generators we also have the commuting generators 
$M_i^{(n)}$ which function like generators of a global symmetry. We can, for instance, consistently set these generators to zero. The presence of these generators therefore do not spoil the ability of the Virasoro-Kac-Moody generators to admit the usual central terms in their commutators. 

What is the meaning of this infinite dimensional extension? Do these additional vector fields generate 
symmetries? 

There is a relatively simple interpretation for the generators $M_i^{(n)}, L^{(n)}, J_a^{(n)}$. 
We know that $P_i=M_i^{(-1)}, B_i=M_i^{(0)}, K_i=M_i^{(1)}$ generate uniform spatial translations, velocity boosts and accelerations respectively.  In fact, it is simple to see from \eq{gcavec} that
the $M_i^{(n)}$ generate arbitrary time dependent (but spatially independent) accelerations. 
\be{acc}
x_i\rightarrow x_i+b_i(t).
\ee
Similarly the $J_{ij}^{(n)}$ in \eq{Jn} generate arbitrary time dependent rotations (once again space independent) 
\be{arbrot}
x_i \rightarrow R_{ij}(t)x_j
\ee
These two set of generators together generate what is sometimes called the Coriolis group: the biggest group of "isometries" of "flat" Galilean spacetime \cite{Duval:1993pe}. 

Recall that in the absence of gravity Galilean spacetime is characterised by a degenerate metric. 
The time intervals are much larger than any space-like intervals in the nonrelativistic scaling limit
\eq{nrelscal}.  We thus have an absolute time $t$ and spatial sections with a flat Euclidean metric. 
We can, in a precise sense, describe the analogue of the isometries  in this Galilean spacetime. The Coriolis group by virtue of preserving the spatial slices (at any given time) are the maximal set of isometries. See appendix A for more details. This realisation of the current algebra in our context is a bit like the occurence of a loop group. 

The generators $L^{(n)}$ have a more interesting action in acting both on time as well as space.
We can read this off from \eq{gcavec}
\be{lnact}
t\rightarrow f(t), \qquad x_i \rightarrow {df\over dt}x_i.
\ee   
Thus it amounts to a reparametrisation of the absolute time $t$. Under this reparametrisation the 
spatial coordinates $x_i$ act as vectors (on the worldline $t$). It seems as if this is some kind of "conformal isometry" of the Galilean spacetime, rescaling coordinates by the arbitrary time dependent 
factor ${df\over dt}$. 

With this interpretation of the infinite extension of the GCA, one might expect that it ought to be partially or fully dynamically realised in physical systems where the finite GCA is (partially or fully) realised.
We will see below an example which lends support to this idea. We will also see in Sec. 5 that  the bulk geometry which we propose as the dual has the extended GCA among its {\it asymptotic} isometries.
An analogy might be two dimensional conformal invariance where the Virasoro algebra is often a symmetry when the finite conformal symmetry of $SL(2,C)$ is realised. And the (two copies of the) Virasoro generators are reflected in the bulk $AdS_3$ as asymptotic isometries. 

Given that the Galilean limit can be obtained by taking a definite scaling limit within a relativistic  theory, we expect to see the GCA (and perhaps its extension) as a symmetry of some subsector within every relativistic conformal field theory. For instance, in the best studied case of ${\cal N}=4$ Yang-Mills 
theory, we ought to be able to isolate a sector with this symmetry. One clue is the presence of the $SL(2,R)$ symmetry together with the preservation of spatial rotational invariance. One might naively think this should be via some kind of conformal quantum mechanics obtained by considering only the spatially independent modes of the field theory. But this is probably not totally correct for the indirect reasons explained in the next paragraph.

Recently, the nonrelativistic limit of the relativistic conformal hydrodynamics, which describes the small fluctuations from thermal equilibrium, have been studied \cite{Fouxon:2008tb, Bhatta:2008kq, Fouxon:2008ik}. One recovers the non-relativistic incompressible Navier-Stokes equation in this limit. The symmetries of this equation were then studied by \cite{Bhatta:2008kq} (see also \cite{Fouxon:2008ik}). One finds that all the generators of the finite GCA are indeed symmetries\footnote{For a realisation of the Schrodinger symmetry in the context of the Navier-Stokes equation see \cite{Hassaine:1999hn, O'Raifeartaigh:2000mp}.} except for the dilatation operator $D$ \footnote{The generator $K$ acts trivially.}.
In particular it has the $K_i$ as symmetries. It is not surprising that the choice of a temperature should break the scaling symmetry of $D$ {\footnote{However, one can define an action of the $\tilde{D}$ as in 
\eq{tildact} to be a symmetry.}}. The interesting point is that the arbitrary accelerations $M_i^{(n)}$ are also actually a symmetry \cite{Russian} (generating what is sometimes called the Milne group \cite{Duval:1993pe}). Thus we have a part of the extended GCA as a symmetry of the non-relativistic Navier-stokes equation which should presumably describe the hydrodynamics in every nonrelativistic field theory. 
In particular, the closed non-relativistic subsector within every relativistic conformal field theory should have a hydrodynamic description governed by the Navier-Stokes equation. This might seem to suggest that this sector ought to have more than just the degrees of freedom of a conformal quantum 
mechanics. 

Coming back to the Navier-Stokes equation, 
if the viscosity is set to zero, one gets the incompressible Euler equations 
\be{eulereqn}
\p_tv_i(x,t)+v_j\p_jv_i(x,t) = -\p_ip(x,t)
\ee
In this case one has the entire finite dimensional GCA being a symmetry since $D$ 
is now also a symmetry.
It is the viscous term which breaks the symmetry under equal scaling of space and time. This shows that one can readily realise "gapless" non-relativistic systems in which space and time scale in the same way! \footnote{Inonu and Wigner \cite{Inonu} have considered representations of the Galilean group without the mass extension and concluded that a particle interpretation of states of the irreducible representations is subtle. In particular such states are not localisable. Just as in the case of relativistic conformal group it is likely that observables such as the S-matrix are ill-defined. We thank Sean Hartnoll for bringing this reference to our attention.}

\section{The Bulk Dual}

Given a particular instance of an AdS/CFT duality, we should be able parametrically to take 
the non-relativistic scaling limit on both sides of the duality. On the field theory side, as we have discussed, the relativistic conformal invariance reduces to the GCA with a possible infinite dimensional dynamical extension. On the string theory side it should be possible to take a similar scaling limit along the lines of the non-relativistic limit studied in \cite{Gomis:2000bd, Danielsson:2000mu, Gomis:2005pg}.
Below we will only consider features of this scaling limit when the parent bulk theory is well described by gravity. This will already involve some novel features. This has to with the fact that 
the usual pseudo-riemannian metric degenerates when one takes a non-relativistic limit. Nevertheless,
there is a well defined, albeit somewhat unfamiliar, geometric description of gravity in such a limit \cite{Misner:1974qy}.
In the (asymptotically) flat space case this is known as the Newton-Cartan theory of gravity which captures Newtonian gravity in a geometric setting. This is a {\it non-metric} gravitational theory. The dynamical variables are affine connections. Einstein's equations reduce to equations for the curvature 
of these non-metric connections. One can generalise this to the case of a negative cosmological constant as well. A variant of this is what we propose below as the right framework for the gravity dual of systems with the GCA. In the next subsection we will briefly review features of the Newton-Cartan theory and then go onto describe the case with a negative cosmological constant. 

\subsection{Newton-Cartan Theory of Gravity}

In the Newton-Cartan description of gravity, the $(d+1)$ dimensional spacetime $M$ has 
a time function $t$ on it which foliates the spacetime into $d$ dimensional spatial slices.  
Stated more precisely (see for example \cite{Ruede:1996sy}): one defines a contravariant tensor $\gamma=\gamma^{\mu \nu}\p_{\mu}\otimes\p_{\nu}$ ($\mu,\nu=0\ldots d$) such there is a time 1-form $\tau=\tau_{\mu}dx^{\mu}$ which is orthogonal to $\gamma$ in the sense that
$\gamma^{\mu\nu}\tau_{\mu}=0$. 
The metric $\gamma$, which has three positive eigenvalues and one zero eigenvalue, will be the non-dynamical spatial metric on slices orthogonal to the worldlines defined by $\tau$. There is no metric on the spacetime as a whole. In fact, its geometric structure is that of a fibre bundle with a one dimensional base (time) and the $d$ dimensional spatial slices as fibres.

The dynamics is encoded in a 
torsion free affine connection $\Gamma^{\mu}_{\nu\lambda}$ on $M$. 
We will demand that this connection is compatible with both $\gamma$ and $\tau$ i.e.
\be{conncomp}
\nabla_{\rho}\gamma^{\mu\nu}=0 \qquad \nabla_{\rho}\tau_{\nu}=0.
\ee
This enables one to define a time function $t$ ("absolute time") 
since we have $\tau_{\mu}=\nabla_{\mu}t$. 
Unlike the Christoffel connections which are determined by the spacetime metric in Einstein's theory, this Newton-Cartan connection is not fixed by just these conditions. One has to impose some additional relations. Defining $R^{\mu\nu}_{\lambda\sigma}=\gamma^{\nu\alpha}R^{\mu}_{\lambda\alpha\sigma}$,
one can define a Newtonian connection as one which obeys the additional condition 
$R^{\mu\nu}_{\lambda\sigma}=R^{\nu\mu}_{\sigma\lambda}.$\footnote{In Einstein gravity with the Christoffel connection and $\gamma$ being the {\it nondegenerate} spacetime metric, this relation is identically satisfied but here it has to be imposed additionally.}

In the presence of matter sources specified by a contravariant second rank stress tensor $T^{\mu\nu}$,
which is additionally required to be covariantly conserved $\nabla_{\mu}T^{\mu\nu}=0$, we can write down the field equations 
which determine the connection in terms of the sources. This is best done by introducing a  "time" like metric $g_{\mu\nu}=\tau_{\mu}\tau_{\nu}$ which is orthogonal to the spatial metric $\gamma_{\mu\nu}$.
The field equations are then familiar in form
\be{newtfldeq}
R_{\mu\nu}=8\pi G(T_{\mu\nu}-{1\over 2}g_{\mu\nu}T)
\ee
where $T_{\mu\nu}=g_{\mu\alpha}g_{\nu\beta}T^{\alpha\beta}$ 
and $T=g_{\alpha\beta}T^{\alpha\beta}$.  
Note that to define the Ricci tensor $R_{\mu\nu}$ (unlike the Ricci scalar $R$) 
one does not need a metric, only the affine connection.

When one chooses coordinates such that 
$\gamma=\delta^{ij}\p_i\otimes\p_j$ ($i,j=1\ldots d$), $\tau=dt$, the non-zero components of the Newtonian connection take the form (imposing appropriate boundary conditions at infinity) 
$\Gamma^{i}_{00}=\p_i\Phi$. The field equations then reduce to Poisson's equations 
with $\Phi$ being the Newtonian gravitational potential and the source being the matter density $\rho=T^{00}$. 

This is, of course, an intrinsic characterisation of Newtonian gravity. Not unsurprisingly, this geometric structure has also been shown to arise in the degenerating limit of a usual Einsteinian geometry \cite{Navarro}. Namely, one can study a one parameter ($\epsilon$) family of usual Lorentzian signature metrics, with the non-relativistic limit $\epsilon \rightarrow 0$ leading to a degenerate metric. 
The condition that this limiting geometry be a Newtonian spacetime is satisfied under fairly mild conditions on the $\epsilon$ dependence of the Lorentzian metric (and therefore the associated geometric objects such as the covariant derivative, curvature tensor etc.). This shows that the nonrelativistic scaling limit is a sensible one to take of a generic Einsteinian geometry. 

\subsection{Newtonian limit of Gravity on $AdS_{d+2}$}

We would like to parametrically carry out the non-relativistic scaling on the bulk $AdS_{d+2}$ 
which would capture the physics of the nonrelativistic limit in the $(d+1)$ dimensional 
boundary theory. In the 
next section we will describe the bulk scaling in more detail. Here we will simply motivate 
its qualitative features and give the resulting Newton-Cartan like description of the bulk 
geometry. 

We know that the boundary metric degenerates in the nonrelativistic limit with the $d$ spatial 
directions scaling as $x_i\propto \epsilon$ while $t \propto \epsilon^0$. We expect this feature to be shared by the bulk metric. One expects that the geometry on constant radial sections to have such a scaling. Since the radial direction of the $AdS_{d+2}$ is an additional  dimension, we have to fix its scaling. The radial direction is a measure of the energy scales in the boundary theory via the holographic correspondence. We therefore expect it to also scale like time i.e. as  $\epsilon^0$. This means that in the bulk  the time and radial directions of the metric {\it both} survive when performing the scaling. Together these constitute an $AdS_2$ sitting inside the original $AdS_{d+2}$. 

What this implies for the dynamics is that we should have a Newton-Cartan like description but with the special role of time being replaced by an $AdS_2$. The geometric structure, in analogy with that of the previous section, is that of  a fibre bundle with $AdS_2$ base and the $d$ dimensional spatial slices 
as fibres. 

Accordingly, we will consider a ("spatial") metric $\gamma=\gamma^{\mu \nu}\p_{\mu}\otimes\p_{\nu}$ ($\mu,\nu=0\ldots d+1$) which now has {\it two} zero eigenvalues corresponding to the time and radial directions. (In a canonical choice of coordinates these directions will correspond  to $\mu=0,d+1$). Mathematically the two null eigenvectors will be taken to span the space of left invariant 1-forms of $AdS_2$. These will 
also define the $AdS_2$ metric $g_{\alpha\beta}$ in the usual way (This is the analogue of the time metric defined in the previous subsection). 

We will once again have dynamical, torsion free affine connections $\Gamma^{\mu}_{\nu\lambda}$ which are compatible with both the spatial and $AdS_2$ metrics 
\be{adsconncomp}
\nabla_{\rho}\gamma^{\mu\nu}=0 \qquad \nabla_{\rho}g_{\alpha\beta}=0.
\ee
There will also be Christoffel connections from the $AdS_2$ and spatial metrics which will not be dynamical if we do not allow these metrics, specifically $g_{\mu\nu}$, to fluctuate. 
We will also impose the condition below \eq{conncomp} on the Riemann tensor. 

In standard Poincare coordinates where $\gamma=z^2\delta^{ij}\p_i\otimes\p_j$ ($i,j=1\ldots d$)
and $g_{\alpha\beta}dx^{\alpha}dx^{\beta}={1\over z^2}(dt^2-dz^2)$, the non-zero components of the 
dynamical connection can be taken to be $\Gamma^{i}_{ab}(t,z,x_i)=\p_i\Phi_{ab}(t,z,x_i)$ (with $a,b=0,d+1$). 
There will be Christoffel components from $\gamma$ and $g$ as mentioned above.

The field equations are the expected modification of \eq{newtfldeq}
\be{adsnewtfldeq}
R_{\mu\nu}-\Lambda g_{\mu\nu}=8\pi G(T_{\mu\nu}-{1\over 2}g_{\mu\nu}T)
\ee
where $T_{\mu\nu}=g_{\mu\alpha}g_{\nu\beta}T^{\alpha\beta}$ 
and $T=g_{\alpha\beta}T^{\alpha\beta}$ and $\Lambda$ is the cosmological constant. 
These are dynamical equations for the fields $\Phi_{ab}(t,z, x_i)$ once the stress tensor 
$T^{ab}(t,z, x_i)$ in the $AdS_2$ directions is specified.

\section{GCA in the Bulk}

In this section we will carry out the non-relativistic scaling limit on the $AdS_5$ piece of the bulk. 
We will also do this for the $SO(4,2)$ isometries of $AdS_5$ and obtain the same contracted algebra 
as in Sec.3\footnote{As mentioned earlier, an algebraically equivalent contraction of the bulk isometries was carried out in \cite{Gomis:2005pg}(see also \cite{Brugues:2006yd}). However the actual embedding of this contracted algebra in $AdS$ is not manifestly that of a nonrelativistic CFT on the boundary. In particular, their foliation of the bulk corresponds to a  boundary geometry which is a time dependent $AdS_2\times S^2$. This is natural from the point of view of considering the worldvolume of a half BPS 
string. By considering Poincare coordinates and foliating the bulk in terms of $R^{3,1}$ slices, it is plausible that their limit will be related directly to ours. We thank Jaume Gomis for helpful communication in this regard.}. 
We will then see how the infinite dimensional extension of this algebra is realised in the bulk. They will have the interpretation as being the generators of asymptotic isometries of the bulk Newton-Cartan like geometry described in the previous section. Since asymptotic isometries, under appropriate circumstances, act on the physical hilbert space of the theory, one finds support for the assertion that the infinite extension can be dynamically realised. 

Consider the metric of $AdS_5$ in Poincare coordinates
\be{poinmet} 
ds^2 = {1\over z^{\prime 2}}(\eta_{\mu\nu}dx^{\mu}dx^{\nu}-dz^{\prime 2})={1\over z^{\prime 2}} (dt^{\prime 2} - dz^{\prime 2} -d x_i^2)
\ee 
The nonrelativistic scaling limit that we will be considering is, as motivated in the previous section
\be{adsnrelscal}
t^{\prime} ,z^{\prime} \rightarrow \epsilon^0 t^{\prime} , \epsilon^0 z^{\prime} \qquad   x_i \rightarrow \epsilon^{1} x_i.
\ee

In this limit we see that only the components of the metric in the $(t^{\prime},z^{\prime})$ directions survive to give the metric on an $AdS_2$. The $d$ dimensional spatial slices parametrised by the $x_i$ 
are fibred over this $AdS_2$. The Poincare patch has a horizon at $z^{\prime}=\infty$ and to extend the coordinates beyond this we will choose to follow an infalling null geodesic, in an analogue of the Eddington-Finkelstein coordinates. Therefore define $z=z^{\prime}$ and $t=t^{\prime}+z^{\prime}$. 
In these coordinates
\be{adsefink}
ds^2 = {1\over z^{2}} (-2dt dz+dt^{2})={dt\over z^{2}}(dt-2dz).
\ee 

\subsection{Contraction of the Bulk Isometries}

In the infalling Eddington-Finkelstein coordinates, the Killing vectors of $AdS_5$ read as 
\ben{primekill}
P_i&=& \p_i, \ B_i= (t-z)\p_i -x_i\p_{t} \cr
K_i&=&(t^{2}-2tz-x_j^2)\p_i+2t x_i\p_{t}+2z x_i\p_{z} +2x_ix_j\p_j \cr
J_{ij} &=& -(x_i\p_j-x_j\p_i) \cr
H&=& -\p_{t}, \ D=-t\p_{t}-z\p_{z}-x_i\p_i \cr
K&=& -(t^{2}+x_i^2)\p_{t}-2z(t-z)\p_{z}-2(t-z)x_i\p_i 
\een
Here we have used the same labelling for the bulk generators as on the boundary  to facilitate easy comparison. Some additional details are given in Appendix B.

Carrying out the scaling \eq{adsnrelscal} we obtain the contracted Killing vectors
\ben{contractkill}
P_i&=& \p_i, \ B_i= (t-z)\p_i ,\quad  K_i =(t^{2}-2tz)\p_i, \quad  J_{ij}= - (x_i\p_j-x_j\p_i)\cr
H&=& -\p_{t}, \quad D=-t\p_{t}-z\p_{z}-x_i\p_i , \ \ \  
K= -t^{2}\p_{t}-2(t-z)(z\p_{z}+x_i\p_i)
\een
We see that at the boundary $z=0$ these reduce to the contracted Killing vectors of the relativistic 
conformal algebra. It can also be checked that these obey the same algebra as \eq{galalg} and
\eq{galconalg} or equivalently \eq{gcafinit} after the relabelling of \eq{rename}. 

The interpretation of most of the generators is straightforward. We note that the $H,K,D$ are scalars under the spatial $SO(d-1)$ and generate, as before, an $SL(2,R)$. We identify this as the isometry group of the $AdS_2$ base of our Newton-Cartan theory.  

We can again define an infinite family of vector fields in the bulk
\ben{bulkinfgen}
M_i^{(m)} &=& (t^{m+1}-(m+1)zt^{m})\p_i \cr
J_{ij}^{(n)} &=& -t^n(x_i\p_j-x_j\p_i) \cr
L^{(n)} &=& -t^{n+1}\p_{t}-(n+1)(t^{ n} -nzt^{n-1})(x_i\p_i+z\p_{z})
\een
These vector fields reduce on the boundary to \eq{gcavec} and \eq{Jn}.

It is rather remarkable that these vector fields also 
obey the commutation relations of the Virasoro-Kac-Moody algebra, the same as in the 
boundary theory
\ben{vkmalgads}
[L^{(m)}, L^{(n)}] &=& (m-n)L^{(m+n)} \qquad [L^{(m)}, J_{a}^{(n)}] = -n J_{a}^{(m+n)} \cr
[J_a^{(n)}, J_b^{(m)}]&=& f_{abc}J_c^{(n+m)} \qquad  [L^{(m)}, M_i^{(n)}] =(m-n)M_i^{(m+n)}. 
\een

How do we interpret all these additional vector fields from the point of view of the bulk?
Firstly, notice that the vector fields $M_i^{(n)}$ and $J_a^{(n)}$ only act on the spatial coordinates 
$x_i$ (with $t,z$ dependent coefficents). From the viewpoint of the fibre bundle structure, these 
are simply rotations and translations on the spatial slices which happen to be dependent on time as well as $z$. These are the isometries of the spatial metric $\gamma$ of the previous section.  They are also trivially isometries of the $AdS_2$ metric since they do not act on those coordinates. In general, these transformations will have a non-trivial effect on the dynamical connection coefficient (though trivial action on the non-dynamical christoffel coefficients). This is not unusual since it is only the vacuum configuration of the bulk theory (in which the dynamical connections vanish) which preserves the full symmetry.  

Now we come to the action of the Virasoro generators, $L^{(n)}$ which act non-trivially on all coordinates. We have under its action (with infinitesimal parameter $a_n$)
\ben{lnact}
z\rightarrow \tilde{z} &=& z[1+a_n(n+1)(t^n-nzt^{n-1})] \cr
t\rightarrow \tilde{t}  &=& t[1+a_nt^n] \cr
x_i \rightarrow \tilde{x_i} &=& x_i[ 1+a_n(n+1)(t^n-nzt^{n-1})]. 
\een
In other words,
\ben{lnactdiff}
dz\rightarrow \tilde{dz} &=& dz[1+a_n(n+1)(t^n-nzt^{n-1})] \cr
 &+& za_nn(n+1)t^{n-2}[(t-(n-1)z)dt -tdz] \cr
dt\rightarrow \tilde{dt}  &=& dt[1+(n+1)a_nt^n] \cr
dx_i \rightarrow \tilde{dx_i} &=& dx_i[ 1+a_n(n+1)(t^n-nzt^{n-1})] \cr
&+& n(n+1)a_nx_it^{n-2}[(t-(n-1)z)dt -tdz]. 
\een
To see how this acts on the Newton-Cartan structure, consider first the above action on the original 
Poincare metric on $AdS_5$ but transformed to the Eddington-Finkelstein coordinates \eq{adsefink}. 
Only after that do we take the scaling limit \eq{adsnrelscal}.
We find 
\ben{lnactmet}
ds^2 &=&{1\over z^{2}} (-2dt dz+dt^{2}+dx_i^2) \rightarrow {1\over z^{2}} (-2dt dz+dt^{2}+dx_i^2) \cr
&+& 2n(n^2-1)a_nt^{n-2}dt^2-2{a_nn(n+1)\over z^2}x_idx_i[(t-(n-1)z)dt -tdz].
\een
We now see that on taking the scaling limit \eq{adsnrelscal} we have
\be{scalact}
ds^2={1\over z^{2}} (-2dt dz+dt^{2}) \rightarrow {1\over z^{2}} (-2dt dz+dt^{2}
+2n(n^2-1)a_nz^2t^{n-2}dt^2).
\ee
As expected the $SL(2,R)$ subgroup $L^{(0)}, L^{(\pm 1)}$ are exact isometries. The other $L^{(n)}$
are not exact isometries. However, they are asymptotic isometries
(See \cite{Brown:1986nw} and references therein).  Near the boundary $z=0$ the diffeomorphisms generated by these vector fields leave the metric unchanged upto a factor which has a falloff like $z^2$. 
Thus these do not affect the non-normalizable mode of the metric.

One expects that when the charges for these asymptotic isometries are constructed, then just as in the Brown-Henneaux construction for $AdS_3$ \cite{Brown:1986nw}(and recent generalisations to $AdS_2$ \cite{Hartman:2008dq}), there will actually be a central term due to boundary contributions. Thus the Virasoro algebra will presumably act in a faithful way on the physical Hilbert space. 

We also notice from the action of the $L^{(n)}$ \eq{lnactmet} on the spatial metric that on the slices of constant $t,z$, the action is again an isometry. 
Thus the $L^{(n)}, J_a^{(n)}, M_i^{(N)}$ together generate (asymptotic) isometries of the spatial and $AdS_2$ metrics $\gamma^{ij}$ and $g_{ab}$. Therefore it seems natural to consider the action of 
these generators on the Newton-Cartan like geometry

\section{Concluding Remarks}

We have seen that the nonrelativistic conformal symmetry obtained as a scaling limit of the relativistic 
conformal symmetry has several novel features which make it a potentially interesting case for further study. The GCA, we have argued, is different from the Schrodinger group which has been studied recently. It also has the advantage of being embedded within the relativistic theory. Hence we ought to have realisations of the GCA in every interacting relativistic conformal field theory. The obvious question is to understand this sector in a particular case such as ${\cal N}=4$ Super Yang-Mills theory. 
And to see whether the infinite dimensional extension can be dynamically realised (and its central charge computed). We have provided indications why this might be the case generically. 

Relatedly, in the bulk gravity dual to such a system
one ought to be able to independently 
compute the central term in the Virasoro algebra {\it a la} Brown-Henneaux. 
In such cases one should be able to use the more general Kac-Moody algebra to constrain the theory and its correlation functions much more. A straightforward generalisation of our results would be to a 
supersymmetric extension of the Kac-Moody algebra. These and related questions are currently under investigation. 

The bulk description in terms of a Newton-Cartan like geometry is somewhat unfamiliar and it would be good to understand it better. In particular, one needs a precise bulk-boundary dictionary to characterise the duality. At least implicitly this is determined by taking the parametric limit of the relativistic duality. 

Then there is the question of how such non-metric theories lift to string theories. This is something we have not touched upon at all in this note. One might hope to get some guidance from previous studies of nonrelativistic string 
theories, though in all these cases one had additional fields like the two form $B_{\mu\nu}$ turned on which made the sigma model well defined. It is therefore not completely clear how to define a string theory on 
these Newton-Cartan like geometries\footnote{There have been works attempting to quantise nonrelativistic theories of gravity directly in a canonical framework \cite{Kuchar:1980tw,Christian:1997wj}. It would be interesting to see if these can be generalised to the case with negative cosmological constant. This could have interesting implications for gauge-gravity dualities in the non-relativistic setting.}.

In the case of the Schrodinger symmetry the dual gravity theory is proposed to live in {\it two} higher dimensions than the field theory. This also provided the route for embedding the dual geometry
in string theory. It is interesting to ask if there is something analogous in our case, whereby the GCA is realised as a standard isometry of a higher dimensional geometry 
(e.g. $(d+3)$ dimensional for a $(d+1)$ dimensional field theory).  

Coming back to the boundary theory, it is interesting to ask whether there are intrinsically 
non-relativistic realisations of the GCA, perhaps in a real life system. It is encouraging in this context 
that the incompressible Euler equations concretely realise the GCA, providing an example of a gapless non-relativistic system.

{\it{Note added}}: It has been brought to our attention that closely related infinite dimensional algebras have been studied in the context of statistical mechanical systems in \cite{Henkel06}. It would be interesting to study the precise connection as well as the potential realisations in statistical mechanics further.

\subsection*{Acknowledgements}

The authors would like to thank Turbasu Biswas for initial collaboration and discussions. We also 
wish to thank K. Balsubramaniam, S. Banerjee, D. Ghoshal, G. Krishnaswami, 
S. Minwalla, V. P. Nair, K. Narayan, S. G. Rajeev, A. Sen, R. Shankar, V. Sreedhar,  
N. Suryanarayana and more generally the participants of the "New Trends in field Theory" conference, BHU, the ISM08 conference and the Pan-IIT meet, Chennai for various questions and comments. 
The work of one of us (R.G.) was supported by the Swarnajayanthi Fellowship of the DST. We would also like to record our indebtedness to the people of India for their generous support for fundamental enquiries. 

\section*{Appendices}
\appendix

\section{Galilean Isometries}

In the Newton-Cartan spacetime described in sec. 4, we do not have a spacetime metric. 
Therefore the usual notion of an isometry as generated by a Killing vector field of the metric is 
not applicable. There is consequently some ambiguity in the definition of an isometry. We will 
paraphrase here some of the different possibilities as outlined in \cite{Duval:1993pe}. 

\begin{enumerate}
\item
{\bf Galilei Algebra}: This consists of all vector fields $X$ satisfying
\be{galcond}
L_{X}\gamma^{\mu\nu}=0 \qquad L_{X}\tau=0 \qquad L_X\Gamma_{\mu\nu}^{\alpha}=0.
\ee 
This gives rise to the usual set of vector fields which generate the finite dimensional Galilean algebra of uniform translations (in space and time), uniform velocity boosts and spatial rotations. 
\item
{\bf Milne Algebra}: This consists of all vector fields $X$ satisfying
\be{milcond}
L_{X}\gamma^{\mu\nu}=0 \qquad L_{X}\tau=0 \qquad L_X\Gamma_{\mu}^{\nu\alpha}=0
\ee 
where $\Gamma_{\mu}^{\nu\alpha}=\gamma^{\beta\nu}\Gamma_{\mu\beta}^{\alpha}$.
The set of vector fields $X$ satisfying this condition is an infinite dimensional extension of the Galilean algebra, now involving arbitrary time dependent boosts/accelerations.
\item
{\bf Coriolis Algebra}: This consists of all vector fields $X$ satisfying
\be{corcond}
L_{X}\gamma^{\mu\nu}=0 \qquad L_{X}\tau=0 \qquad L_X\Gamma^{\mu\nu\alpha}=0
\ee 
where $\Gamma^{\mu\nu\alpha}=\gamma^{\rho\mu}\gamma^{\beta\nu}\Gamma_{\rho\beta}^{\alpha}$.
The set of vector fields $X$ satisfying this condition is a further infinite dimensional extension of the Milne algebra, now involving in addition to the arbitrary boosts or accelerations, arbitrary time dependent rotations as well.
\end{enumerate}

\bigskip

\section{Killing vectors of $AdS_{d+2}$ and Bulk Contraction}

Here we list the killing vectors of $AdS_{d+2}$ in the $d+3$ dimensional Minkowskian embedding space and then rewrite them in intrinsic AdS coordinates. For taking the contraction, we will work in Poincare coordinates. 

We denote flat $d+1$-dimensional space with co-ordinates $y_a$ with $a=1,..,d+1$.
Embedding equation:
\ben{embed}
uv + \eta_{ab} y^a y^b &=& 1 \\
ds^2 &=& du dv + \eta_{ab} dy^a dy^b
\een
where $\eta_{ab}$ has a form $diag (1, -1, -1, \ldots -1)$. Also note that we have rescaled the $AdS_{d+2}$ radius to $1$.

We wish to write in explicit $SO(d+1,2)$ notation. So we choose $u= y_0 + y_{d+2}$ 
and $v=y_0 -y_{d+2}$. 
Now we would be interested in the co-ordinates on the Poincare patch. 
\ben{patch}
z &=& v^{-1} = (y_0 - y_{d+2})^{-1} \\ 
t &=& v^{-1} y_1 = (y_0 - y_{d+2})^{-1} y_1\\
x_i &=& v^{-1} y_i = (y_0 - y_{d+2})^{-1} y_i 
\een 
where we label $i = 2,3\ldots d+1$.

The constraint equation becomes:
\be{}
u = z - (z)^{-1} (t^2 - x_i^2)
\ee
which gives the metric on the Poincare patch \refb{poinmet}. 

We can take the inverse transformations and express the derivatives of the Poincare co-ordinates in terms of derivatives of $y$'s and take various
linear combinations to obtain 
\begin{eqnarray*} \label{PP-killing}
M_{01} &=& -(y_0 \p_1 - y_1 \p_0) \\ 
&=&  -{1\over 2} (z^2 + 1 + t^2 + x_i^2) {\p \over {\p t}} - z t {\p \over {\p z}} - t x_i {\p \over {\p x_i}} \\
M_{0i} &=& -(y_0 \p_i + y_i \p_0 )\\ 
&=&  -{1\over 2} (z^2 + 1 - t^2 + +x_i^2 +x_j^2) {\p \over {\p x_i}} +z x_i {\p \over {\p z}} + t x_i {\p \over {\p t}} +x_i^2{\p \over {\p x_i}} +x_i x_j {\p \over {\p x_j}} \quad {\mbox{($j\neq i$)}} \\ 
M_{1i} &=& -(y_1 \p_i + y_i \p_1) =  -t {\p \over {\p x_i}} - x_i {\p \over {\p t}} \\
M_{ij} &=& -(y_i \p_j - y_j \p_i) = -(x_i {\p \over {\p x_j}} - x_j {\p \over {\p x_i}}) \\
M_{0,d+2} &=& -(y_0 \p_{d+2} + y_{d+2} \p_0) = -z {\p \over {\p z}} -t {\p \over {\p t}}- x_i {\p \over {\p x_i}} \\
M_{1,d+2} &=& -(y_1 \p_{d+2} + y_{d+2} \p_1)  \\ 
&=& -{1\over 2} (z^2 + t^2 + x_i^2 -1) {\p \over {\p t}} -z t {\p \over {\p z}} -t x_i {\p \over {\p x_i}} \\
M_{i,d+2} &=& -(y_{d+2} \p_i + y_i \p_{d+2})  \\ 
&=&  -{1\over 2} (z^2 - t^2 + x_i^2 + x_j^2- 1) {\p \over {\p x_i}} + z x_i {\p \over {\p z}} + t x_i {\p \over {\p t}} + x_i^2{\p \over {\p x_i}} + x_i x_j {\p \over {\p x_j}} \quad {\mbox{($j\neq i$)}} 
\end{eqnarray*}
In the above equations, the repeated indices $j$ are summed over in $M_{0i}$ and $M_{i,d+2}$.

To connect with our notation for the boundary generators, we define:
\begin{eqnarray}
H &=&  M_{01} -  M_{1,d+2} ; \quad  K =  M_{01} +  M_{1,d+2}; \quad D =  M_{0,d+2} \nonumber \\
P_i &=&  -M_{0i} +  M_{i,d+2}; \quad K_i =  M_{0i} +  M_{i,d+2}; \quad B_i = - M_{1i} \\
M_{ij} &=&  M_{ij} \nonumber
\end{eqnarray}

After transforming to infalling Eddington-Finkelstein coordinates the generators of \eq{PP-killing}
become then ones given in  \eq{primekill}
We then carry out the contraction on these Killing vectors using the scaling:
$z = \e^0 \tilde z$, $t = \e^0 \tilde t$ and $x_i = \e^1 \tilde x_i$ and obtain \refb{contractkill}.


\end{document}